\documentclass[showpacs,preprintnumbers,amsmath,amssymb,APSl,prd,nofootinbib,superscriptaddress, onecolumn]{revtex4}
\usepackage[dvips]{graphicx}
\usepackage{amssymb}
\usepackage{amsmath}
\usepackage{graphicx,color}
\usepackage{amsfonts}
\usepackage{bm}
\usepackage{cancel}
\usepackage{comment}
\usepackage{hyperref}

\usepackage{babel}

\newcommand{\be}{\begin{equation}}
\newcommand{\ee}{\end{equation}}
\newcommand{\bea}{\begin{eqnarray}}
\newcommand{\eea}{\end{eqnarray}}
\newcommand{\beaa}{\begin{eqnarray*}}
\newcommand{\eeaa}{\end{eqnarray*}}

\newcommand{\e}{\mathrm{e}}

\allowdisplaybreaks[4]

\begin{document}

\tolerance=5000

\title{Unifying inflation with early and late dark energy in Einstein-Gauss-Bonnet gravity}

\author{Shin'ichi Nojiri}
\email{nojiri@gravity.phys.nagoya-u.ac.jp}
\affiliation{Department of Physics, Nagoya University, Nagoya 464-8602, Japan}
\affiliation{Kobayashi-Maskawa Institute for the Origin of Particles and the Universe, Nagoya University, Nagoya 464-8602, Japan}
\author{Sergei~D.~Odintsov}
\email{odintsov@ice.csic.es} \affiliation{Institute of Space Sciences (ICE, CSIC) C. Can Magrans s/n, 08193 Barcelona, Spain }
\affiliation{Instituci\'o Catalana de Recerca i Estudis Avan\c{c}ats (ICREA),
Passeig Luis Companys, 23, 08010 Barcelona, Spain}
\author{Diego~S\'aez-Chill\'on~G\'omez}
\email{diego.saez@uva.es} \affiliation{Department of Theoretical, Atomic and Optical
Physics, Campus Miguel Delibes, \\ University of Valladolid UVA, Paseo Bel\'en, 7, 47011
Valladolid, Spain}
 \affiliation{Mathematics Research Institute IMUVA, Campus Miguel Delibes, Edificio LUCIA, \\ University of Valladolid UVA, Paseo Bel\'en, S/N, 47011
Valladolid, Spain}
\begin{abstract}
In the era of precision cosmology, different observational data has led to precise measurements of the Hubble constant that differ significantly, what has been called the Hubble tension problem. In order to solve such a discrepancy, many different solutions have been proposed, from systematic errors on the observational data to theoretical proposals that assume an early dark energy that might affect the universe expansion at the time of recombination. In this paper, a model of varying cosmological constant is proposed in the framework of Einstein-Gauss-Bonnet gravity. The corresponding gravitational action is reconstructed and such a model is shown to reproduce well the inflationary era together with dark energy epoch and at the same time to provide an explanation for the discrepancy on the Hubble constant predictions. The transition to a phantom epoch is also realised, avoiding the usual instability problems of ordinary scalar field models.
\end{abstract}

\maketitle

\section{introduction}

As pointed out by different sources of observations, there exists a conflict on the predicted Hubble constant value from observational data. While the Planck collaboration through observations of the Cosmic Microwave Background (CMB) and 
the posterior data analysis shed a value of the Hubble constant given by $H_0 = 67.36 \pm 0.54$ km/s/Mpc~\cite{Planck:2018vyg}, 
some other observations based on late-time sources provide a greater value. 
For instance, the Supernovae $H_0$ for the Equation of State (SH0ES) collaboration estimates $H_0 = 73.5 \pm 1.4$ km/s/Mpc~\cite{Riess:2020fzl}. Similarly other low-$z$ measurements, as H0LiCOW returns a value of $H_0=73.3^{+1.7}_{-1.8}$ km/s/Mpc~\cite{Pesce:2020xfe} 
while the Megamaser Cosmology Project predicts $H_0=73.9 \pm 3.0$ km/s/Mpc~\cite{Reid:2012hm,Kuo:2012hg,Gao:2015tqd}. 
Such discrepancy reaches a $4\sigma$ tension in some cases. 
This problem is the so-called ``Hubble Tension'' problem~\cite{Verde:2019ivm}. While such tension might be a consequence of systematic errors in the measurements, 
the existing tension among different sources of data would imply the same errors over all of them. 
Hence, over the last few years, some other ways to solve such a problem have been explored theoretically. 
One of them is based on a hypothetical variation of the cosmological constant along the universe expansion, 
such that the value during recombination does not remain constant till the current epoch and consequently the analysis by the Planck collaboration 
mislead such variation that would affect the resulting Hubble constant. 
This hypothesis can be modelled in many different ways but all of them have been referred to as ``Early Dark Energy''~\cite{Poulin:2018cxd, 
Mortsell:2018mfj, Niedermann:2020dwg, DiValentino:2017iww, Yang:2018euj, Pan:2019jqh, Gomez-Valent:2020mqn, Pan:2020zza, DEramo:2018vss, 
Vagnozzi:2019ezj, Ye:2020btb, Nunes:2018xbm, Wang:2020zfv, Odintsov:2020qzd, Nojiri:2021dze}. 
In this scenario, the cosmological constant is not constant but varies in time, such that it evolutes and changes its value in the early universe. 
For early universe observations, the estimation of the Hubble constant is based on angular scales and consequently on the ratio of the physical scale 
and the angular distance to the CMB, such that if they are modified, a different value of the Hubble constant is obtained. 
To achieve a varying cosmological constant, there have been some attempts by using modified gravity theories like $F(R)$ 
gravity as in~\cite{Nojiri:2022ski} (for a general review on modified gravity theories, 
see~\cite{Capozziello:2009nq, Faraoni:2010pgm, Capozziello:2011et, Nojiri:2010wj, Nojiri:2017ncd}). 

As far as the early cosmological constant was smaller than the present one, there might be a period during which the Hubble rate $H$ increases, $\dot H>0$. 
In Einstein's gravity, for a flat Friedmann-Lema\^itre-Robertson-Walker (FLRW) spacetime, the corresponding equations yield: 
\begin{align}
\label{FRWeqs}
\frac{3}{\kappa^2} H^2 = \rho\, , \quad - \frac{1}{\kappa^2} \left( 3H^2 + 2 \dot H \right)=p\, ,
\end{align}
where $\rho$ and $p$ are the energy density and the pressure of matter respectively, such that the equation of state (EoS) parameter is given by: 
\begin{align}
\label{totalw1}
w_\mathrm{total}=\frac{p}{\rho}=- 1 - \frac{2\dot H}{3H^2}\, .
\end{align}
Hence, as far as $\dot H>0$, the EoS parameter is $w_\mathrm{total} \leq -1$. 
Such types of fluids are called phantom~\cite{Caldwell:1999ew} and violate the null energy condition. 
Therefore, an effective phantom fluid is necessary to solve the Hubble tension problem in this way. 
Consequently, a natural scenario to realize such a mechanism might be described by phantom dark energy, which can mimic a varying cosmological constant. 
In the phantom dark energy scenario, the energy density increases as the expansion of the universe does, and the effective cosmological constant increases with time too. 
In~\cite{Gangopadhyay:2022bsh}, such a scenario is proposed where the coupling between the baryonic matter and the dark matter produces a phantom.  
This transition can be also realized by using a scalar field~\cite{Nojiri:2005pu} but to have a stable phantom-crossing barrier, 
several scalar fields are required~\cite{Capozziello:2005tf,Elizalde:2008yf}. 
Furthermore, in a phantom universe within simple scalar-tensor theory, the scalar field becomes a ghost, which induces instabilities at the quantum level. 

In this paper, we focus on the reconstruction of a varying cosmological constant model within scalar-Einstein-Gauss-Bonnet gravity (EGB), 
where a phantom transition might be achieved without instabilities~\cite{Nojiri:2005vv}. Other scenarios where a varying cosmological constant is proposed have been analysed in the literature \cite{Cabral:2020mst,Guendelman:2013sca}. In EGB gravity, the scalar field $\phi$ is coupled to the Gauss-Bonnet (GB) invariant, motivated by stringy corrections~\cite{Nojiri:2006je,Cognola:2006}. 
Within these models, one might describe correctly the whole cosmological evolution by an effective varying cosmological constant 
whose evolution reproduces the inflationary period in the very early universe, and later the dark energy epoch in the present universe, 
while its evolution also explains the differences in the measurements of the Hubble constant among the early and late-time observational data. 
EGB gravity has been analyzed previously in the literature, showing a successful realization of inflation 
within these models~\cite{Guo:2009uk, Jiang:2013gza, KohLLT:2014, Kanti:2015pda, Yi:2018gse, Odintsov:2018zhw, NojiriOOCP:2019, 
Odintsov:2020sqy, Odintsov:2020mkz, OikonomouF:2020nm, Fomin:2020, PozdeevaGSTV:2020,Oikonomou:2022xoq,Oikonomou:2021kql,Oikonomou:2015qha,Hwang:2005hb}. In addition, black holes have been also analysed in EGB gravities, showing some different solutions in comparison with GR \cite{Rubiera-Garcia:2015yga,Tangphati:2021tcy,Hansraj:2020rvc,Papnoi:2021rvw,Charmousis:2021npl,EslamPanah:2020hoj,Zhang:2020sjh}
Also, models that reproduce the late-time acceleration have been considered in the framework of the Gauss-Bonnet gravities~\cite{Nojiri:2005vv, Cognola:2006}, 
which show that the models can be compatible with observational data~\cite{Odintsov:2022zrj}. 

Here we focus on the realization of a varying cosmological constant within EGB gravity that might reproduce inflation and late-time acceleration, 
and alleviate the Hubble tension.To do so, we propose a particular model that can be read as an effective cosmological constant whose value is evolving with the cosmological expansion, such that during each different epoch it falls into an effective cosmological constant and then transits to the next stage. All this mechanism is shown to be reproduced within EGB gravity, since it provides a simple way to recast the corresponding cosmological behaviour and to obtain the gravitational action. We show that Early Dark Energy can be described by such model and also a phantom transition is achieved.

The paper is organized as follows: In Section~\ref{Sec1} EGB gravity is introduced. 
Section~\ref{SecII} is devoted to the reconstruction procedure for obtaining a unified model. 
Inflation is analyzed in Section~\ref{SecIII} while Phantom dark energy is studied in Section~\ref{SecIV}. 
Finally, in Section~\ref{SecV} we introduce a generalization of EGB gravity and the corresponding equations.

\section{scalar-Einstein-Gauss-Bonnet gravity in FLRW spacetime}\label{Sec1}

Let us start by introducing the gravitational action for the so-called scalar-Einstein-Gauss-Bonnet gravity:
\begin{align}
\label{SEGB1}
S=\int d^4 x \sqrt{-g}\left[ \frac{R}{2\kappa^2} 
 - \frac{1}{2}\partial_\mu \phi \partial^\mu \phi 
 - V(\phi) - \xi(\phi) \mathcal{G}\right]\, , 
\end{align}
where the Gauss-Bonnet invariant $\mathcal{G}$ is given by:
\begin{align}
\mathcal{G} \equiv R^2 - 4 R_{\mu\nu} R^{\mu\nu} + R_{\mu\nu\rho\sigma} R^{\mu\nu\rho\sigma}\, .
\label{GBterm}
\end{align}
Note that as far as one sets the coupling $\xi(\phi)$ to be constant, the Gauss-Bonnet term does not contribute to the field equations because $\mathcal{G}$ is a topological invariant. 
The corresponding field equations are obtained by varying the action (\ref{SEGB1}) with respect to the metric and to the scalar field, which leads to:
\begin{align}
\label{scalarGB1}
0= \Box \phi - V'(\phi) - \xi'(\phi) \mathcal{G}\, .
\end{align}
This is the equation for the scalar field $\phi$, whereas the gravitational field equations are: 
\begin{align}
\label{gb4b}
T^{\mu \nu}= &\, \frac{1}{2\kappa^2}\left(- R^{\mu\nu} + \frac{1}{2} g^{\mu\nu} R\right)
+ \left(\frac{1}{2} \partial^\mu \phi \partial^\nu \phi
 - \frac{1}{4}g^{\mu\nu} \partial_\rho \phi \partial^\rho \phi \right) - \frac{1}{2} g^{\mu\nu} V(\phi) \nonumber \\
&\, + 2 \left( \nabla^\mu \nabla^\nu \xi(\phi)\right)R - 2 g^{\mu\nu} \left( \nabla^2\xi(\phi)\right)R
 - 4 \left( \nabla_\rho \nabla^\mu \xi(\phi)\right)R^{\nu\rho}
 - 4 \left( \nabla_\rho \nabla^\nu \xi(\phi)\right)R^{\mu\rho} \nonumber \\
&\, + 4 \left( \nabla^2 \xi(\phi) \right)R^{\mu\nu}
+ 4g^{\mu\nu} \left( \nabla_{\rho} \nabla_\sigma \xi(\phi) \right) R^{\rho\sigma} 
 - 4 \left(\nabla_\rho \nabla_\sigma \xi(\phi) \right) R^{\mu\rho\nu\sigma}\, .
\end{align}
Here $T^{\mu \nu}$ is the energy-momentum tensor for the matter fields. 
We should note that the field equations~(\ref{gb4b}) remain second order 
since no derivatives of curvature arise in the equations due to the linearity of the action with respect to the curvature and the Gauss-Bonnet term. 
Along this paper, we are considering a flat FLRW metric: 
\begin{align}
\label{FRW}
ds^2= -dt^2 + a(t)^2\sum_{i=1,2,3} \left(dx^i\right)^2\, .
\end{align} 
Here the metric is expressed in comoving coordinates, where $t$ is the cosmological time and $a(t)$ is the scale factor. 
Hence, the $(00)$ and $(ij)$ components of the field equations~(\ref{gb4b}) turn out:
\begin{align}
\label{SEGB3}
0=&\, - \frac{3}{\kappa^2}H^2 + \frac{1}{2}{\dot\phi}^2 + V(\phi) + 24 H^3 \frac{d \xi(\phi(t))}{dt} + \rho \, ,\nonumber \\
0=&\, \frac{1}{\kappa^2}\left(2\dot H + 3 H^2 \right) + \frac{1}{2}{\dot\phi}^2 - V(\phi) 
 - 8H^2 \frac{d^2 \xi(\phi(t))}{dt^2} - 16H \dot H \frac{d\xi(\phi(t))}{dt} - 16 H^3 \frac{d \xi (\phi(t))}{dt}+ p\, ,
\end{align}
where $H\equiv \frac{\dot a}{a}$ is the Hubble parameter. In addition, $\rho$ and $p$ are the energy density and pressure of the matter fields, again, respectively, 
that gather all the matter contents:
\begin{align}
\label{SGBEG9_0}
\rho = \sum_m \rho_m \, , \quad p = \sum_m p_m = \sum w_m \rho_m \, , \quad w_m \equiv \frac{p_m}{\rho_m} \, ,
\end{align}
each matter component satisfying the continuity equation:
\begin{align}
\label{SGBEG9}
0 = \dot \rho_m + 3H \left( \rho_m + p_m \right) \, ,
\end{align}
which can be easily solved in terms of the scale factor, leading to: 
\begin{align}
\label{matters}
\rho_m = \rho_{m0} a^{-3 \left( 1 + w_m \right)} 
= \rho_{m0} \e^{-3 \left( 1 + w_m \right) N}\, .
\end{align}
Here $\rho_{m0}$ is a constant and $N$ is the number of $e$-foldings $a=\e^N$. 
For convenience, the field equations~(\ref{SEGB3}) can be expressed in terms of $N$ instead of the cosmic time $t$ as the independent variable:
\begin{align}
\label{SEGB3N}
0=&\, - \frac{3}{\kappa^2}H^2 + \frac{1}{2}H^2{\phi'}^2 + V(\phi) + 24 H^4 \frac{d \xi(\phi(N))}{dN} 
+ \sum_m \rho_{m0} \e^{-3 \left( 1 + w_m \right) N} \, ,\nonumber \\
0 =&\, \frac{1}{\kappa^2}\left(2H H' + 3 H^2 \right) + \frac{1}{2}H^2 {\phi'}^2 - V(\phi) 
 - 8 \left( 2 H^4 \frac{d \xi (\phi(N))}{dN} + 3 H^3 H' \frac{d \xi(\phi(N))}{dN} + H^4 \frac{d^2 \xi(\phi(N))}{dN^2} \right) \nonumber \\
&\, + \sum_m w_m \rho_{m0} \e^{-3 \left( 1 + w_m \right) N} \, ,
\end{align}
where $'=\frac{d}{dN}$ and we have used $\frac{d\phi}{dt}=H\frac{d\phi}{dN}$, etc. In the case of Einstein's gravity (with a cosmological constant $\Lambda$), 
the first FLRW equation reduces to:
\begin{align}
\label{GBEDE6_0}
H(N)^2=\frac{\kappa^2}{3} \left( \Lambda + \sum_m \rho_{m0} \e^{-3 \left( 1 + w_m \right) N} \right) \, .
\end{align}
In this paper, we intend to describe inflation, early dark energy, and dark energy under the same model, such that it might be seen 
as a standard $\Lambda$CDM model with a varying cosmological constant:
\begin{align}
\label{GBEDE7_0}
H(N)^2=\frac{\kappa^2}{3} \left( \Lambda (N) + \sum_m \rho_{m0} \e^{-3 \left( 1 + w_m \right) N} \right) \, .
\end{align}
By comparing first equation in (\ref{SEGB3N}) with (\ref{GBEDE7_0}), we find:
\begin{align}
\label{SEGB16}
\Lambda(N) = \frac{1}{2} H^2 {\phi'}^2 + V(\phi) + 24 H^4 \frac{d \xi(\phi)}{d\phi} \phi' \, .
\end{align}
In addition, from the scalar field equation~(\ref{scalarGB1}), we can define an effective potential $V_\mathrm{eff}(\phi)$ for $\phi$ given by:
\begin{align}
\label{SEGB12_0}
V_\mathrm{eff}(\phi) = V(\phi) + \xi(\phi) \mathcal{G} = V(\phi) + 24 H^3 \left( H' + H \right) \xi(\phi) \, ,
\end{align}
Hence, the field equation~(\ref{scalarGB1}) for $\phi$ in the FLRW spacetime (\ref{FRW}) has the following form, 
\begin{align}
\label{SEGB15}
0= \ddot \phi + 3 H \dot \phi + \frac{dV_\mathrm{eff}(\phi)}{d\phi} \, .
\end{align}
By this equation~(\ref{SEGB15}), one can note that the dynamics of the scalar field $\phi$ is governed by the effective potential $V_\mathrm{eff}(\phi)$. 
Then, if the scalar field is on an extremum of $V_\mathrm{eff}(\phi)$, it may stay there, where $\phi'=\dot\phi=0$. 
In such a case, the varying cosmological constant becomes $\Lambda(N) = V(\phi)$ but is not given by $V_\mathrm{eff}(\phi)$ 
despite the scalar field might not be on an extremum of $V(\phi)$.
We should note that the effective potential $V_\mathrm{eff}(\phi)$ depends on the Hubble rate and its first derivative, or in other words, on the spacetime curvature. 
Therefore the minimum of the effective potential $V_\mathrm{eff}(\phi)$ changes according to the change in curvature 
and consequently, a phase transition might occur along the cosmological evolution, in analogy to an effective potential at a finite temperature, 
where a phase transition might occur depending on the temperature.

\section{Reconstructing a unified model that includes early Dark Energy}\label{SecII}

In this section, firstly we introduce the reconstruction procedure based on the one developed in Refs.~\cite{Nojiri:2006je} and~\cite{Nojiri:2009kx}, 
and then, we propose a model that might realize inflation, early dark energy, and dark energy under the same mechanism. 
The scalar potential can be dropped out by combining both equations in (\ref{SEGB3N}) leading to:
\begin{align}
\label{GBEDE1}
0=&\, \frac{2H H'}{\kappa^2} + H^2 {\phi'}^2 
 - 8 H \e^N \frac{d}{dN} \left( \e^{-N} H^3 \frac{d \xi (\phi(N))}{dN} \right) + \sum_m \left( 1 + w_m\right) \rho_{m0} \e^{-3 \left( 1 + w_m \right) N} \, ,
\end{align}
which can be integrated to obtain $\xi\left(\phi\left(N\right)\right)$ as follows: 
\begin{align}
\label{GBEDE2}
\xi\left(\phi\left(N\right)\right) =&\, \frac{1}{8} \int_{N^{(2)}}^N dN_1\frac{\e^{N_1}}{H \left( N_1 \right)^3} \int_{N^{(1)}}^{N_1} dN_2 \e^{-N_2} \nonumber \\
&\, \times \left\{ \frac{2H' \left(N_2 \right)}{\kappa^2} + H\left( N_2 \right) \phi' \left( N_2 \right)^2 
+ \frac{1}{H\left(N_2\right)} \sum_m \left( 1 + w_m\right) \rho_{m0} \e^{-3 \left( 1 + w_m \right) N_2} \right\}\, .
\end{align}
Here $N^{(1)}$ and $N^{(2)}$ are integration constants. 
Furthermore, by using the first equation in (\ref{SEGB3N}), the expression for the scalar potential is finally achieved:
\begin{align}
\label{GBEDE3}
V\left(\phi\left(N\right)\right) =&\, \frac{3}{\kappa^2}H\left(N\right)^2 - \frac{1}{2}H\left(N\right)^2 \phi'\left(N\right)^2 
 - 3 H\left(N\right) \e^N \int_{N^{(1)}}^N dN_1 \e^{-N_1} \left\{ \frac{2H' \left(N_1 \right)}{\kappa^2} \right. \nonumber \\
&\, \left. + H\left( N_1 \right) \phi' \left( N_1 \right)^2 
+ \frac{1}{H\left(N_1\right)} \sum_m \left( 1 + w_m\right) \rho_{m0} \e^{-3 \left( 1 + w_m \right) N_1} \right\} 
 - \sum_m \rho_{m0} \e^{-3 \left( 1 + w_m \right) N} \, . 
\end{align}
Hence, by assuming a model described by the following scalar potential and coupling in terms of two functions $g=g(N)$ and $f=f(\phi)$:
\begin{align}
\label{GBEDE4}
\xi\left(\phi\right) =&\, \frac{1}{8} \int_{N^{(2)}}^{f(\phi)} dN_1\frac{\e^{N_1}}{g \left( N_1 \right)^3} \int_{N^{(1)}}^{N_1} dN_2 \e^{-N_2} \nonumber \\
&\, \times \left\{ \frac{2g' \left(N_2 \right)}{\kappa^2} + \frac{g \left( N_2 \right)}{f' \left( f^{-1}\left(N_2\right) \right)^2} 
+ \frac{1}{g\left(N_2\right)} \sum_m \left( 1 + w_m\right) \rho_{m0} \e^{-3 \left( 1 + w_m \right) N_2} \right\}\, , \\
\label{GBEDE5}
V\left(\phi\right) =&\, \frac{3}{\kappa^2}g\left(f\left(\phi\right)\right)^2 - \frac{g\left(f\left(\phi\right)\right)^2}{2f'\left(\phi \right)^2} 
 - 3 g\left(f\left(\phi\right)\right) \e^{f(\phi)} \int_{N^{(1)}}^{f(\phi)} dN_1 \e^{-N_1} \left\{ \frac{2 g' \left(N_1 \right)}{\kappa^2} \right. \nonumber \\
&\, \left. + \frac{g\left( N_1 \right)}{f' \left( f^{-1}\left(N_1\right) \right)^2} 
+ \frac{1}{g\left(N_1\right)} \sum_m \left( 1 + w_m\right) \rho_{m0} \e^{-3 \left( 1 + w_m \right) N_1} \right\} 
 - \sum_m \rho_{m0} \e^{-3 \left( 1 + w_m \right) f^{-1}(\phi)} \, .
\end{align}
The corresponding solutions for the Hubble rate and the scalar field are obtained:
\begin{align}
\label{GBEDE6}
H(N)=g\left( N \right)\, , \quad \phi=f^{-1}\left(N\right)\, ,
\end{align}
where $f^{-1}(N)$ is the inverse function of $f(N)$. Note also that by using (\ref{GBEDE7_0}), 
the function $g(N)$ can be expressed in terms of the varying cosmological constant as follows:
\begin{align}
\label{GBEDE7}
g(N)^2=\frac{\kappa^2}{3} \left( \Lambda (N) + \sum_m \rho_{m0} \e^{-3 \left( 1 + w_m \right) N} \right) \, .
\end{align}
Then, the expressions for the scalar potential and coupling (\ref{GBEDE4},\ref{GBEDE5}) can be rewritten as:
\begin{align}
\label{GBEDE8}
\xi\left(\phi\right) =&\, \frac{1}{8} \int_{N^{(2)}}^{f(\phi)} dN_1\frac{\e^{N_1}}{g \left( N_1 \right)^3} \int_{N^{(1)}}^{N_1} dN_2 \frac{\e^{-N_2}}{g\left(N_2\right)} 
\left\{ \frac{\Lambda' \left(N_2\right)}{3} + \frac{g \left( N_2 \right)^2}{f' \left( f^{-1}\left(N_2\right) \right)^2} \right\}\, , \\
\label{GBEDE9}
V\left(\phi\right) =&\, \Lambda \left( f \left(\phi\right) \right) - \frac{g\left(f\left(\phi\right)\right)^2}{2f'\left(\phi \right)^2} 
 - 3 g\left(f\left(\phi\right)\right) \e^{f(\phi)} \int_{N^{(1)}}^{f(\phi)} dN_1 \frac{\e^{-N_1}}{g\left(N_1\right)} \left\{ \frac{\Lambda' \left(N_1\right)}{3} 
+ \frac{g\left( N_1 \right)^2}{f' \left( f^{-1}\left(N_1\right) \right)^2} \right\} \, , 
\end{align}
In addition, there is still a degree of freedom in the choice of $f(\phi)$, such that we may identify the scalar field $\phi$ 
with the number of $e$-foldings $\phi=N$, which simplifies the calculations. Another possible choice for $f(N)$ might be such that:
\begin{align}
\label{GBEDE10}
\frac{1}{g\left(N\right)} \left\{ \frac{\Lambda' \left(N\right)}{3} 
+ \frac{g\left( N \right)^2}{f' \left( f^{-1}\left(N\right) \right)^2} \right\}=\frac{2}{\kappa^3} \, .
\end{align}
By such a choice, the expressions in (\ref{GBEDE8}) and (\ref{GBEDE9}) turns out: 
\begin{align}
\label{GBEDE11}
\xi\left(\phi\right) = \frac{1}{4\kappa^3} \int_{N^{(2)}}^{f(\phi)} \frac{dN_1}{g \left( N_1 \right)^3} \, , \quad 
V\left(\phi\right) = \Lambda \left( f \left(\phi\right) \right) + \frac{\Lambda' \left( f \left(\phi\right) \right)}{6} - \frac{7}{\kappa^3} g\left(f\left(\phi\right)\right) \, , 
\end{align}
where we have suppressed an integration constant by assuming $N^{(1)}\to + \infty$. Then, Eq.~(\ref{GBEDE10}) can be rewritten as follows:
\begin{align}
\label{phi1}
\frac{d\phi}{dN}= \pm \sqrt{ \frac{2}{\kappa^3 g(N)} - \frac{\Lambda'(N)}{3g(N)^2}} \, ,
\end{align}
which can be easily solved, leading to: 
\begin{align}
\label{phi2}
\phi\left(N \right) = \pm \int^N dN'\sqrt{ \frac{2}{\kappa^3 g \left( N' \right)} - \frac{\Lambda'\left( N' \right)}{3g \left( N' \right)^2}} \, .
\end{align}
Hence, by solving (\ref{phi2}) with respect to $N$, we can obtain $N=f(\phi)$ in (\ref{GBEDE6}). 
It is straightforward to show the consistency of all the procedures by using the equations (\ref{GBEDE6}), (\ref{GBEDE8}), and (\ref{GBEDE9}) to obtain the scalar potential:
\begin{align}
\label{SEGB12}
\frac{dV_\mathrm{eff}(\phi)}{d\phi} =&\, - \frac{g\left(f\left(\phi\right)\right)g' \left(f\left(\phi\right)\right)}{f'\left(\phi \right)} 
+ \frac{g\left(f\left(\phi\right)\right)^2f''\left(\phi \right)}{f'\left(\phi \right)^3} - \frac{3 g\left( f\left(\phi\right) \right)^2}{f' \left( \phi \right)} \, .
\end{align}
Since $f'(\phi)=\frac{dN}{d\phi}$ one finds:
\begin{align}
\label{SEGB13} 
\frac{d^2 \phi}{dN^2} = \frac{d}{dN} \left( \frac{1}{f'(\phi)} \right) = \frac{d\phi}{dN} \frac{d}{d\phi} \left( \frac{1}{f'(\phi)} \right) = - \frac{f''(\phi)}{f'(\phi)^3} \, ,
\end{align}
which leads to: 
\begin{align}
\label{SEGB14} 
\ddot \phi + 3 H \dot \phi =&\, H^2 \frac{d^2 \phi}{dN^2} + H H' \frac{d\phi}{dN} + 3 H^2 \frac{d\phi}{dN} \nonumber \\
=&\, - \frac{g\left(f\left(\phi\right)\right)^2f''\left(\phi \right)}{f'\left(\phi \right)^3} + \frac{g\left(f\left(\phi\right)\right)g' \left(f\left(\phi\right)\right)}{f'\left(\phi \right)} 
+ \frac{3 g\left( f\left(\phi\right) \right)^2}{f' \left( \phi \right)} \, .
\end{align}
Hence, Eq.~(\ref{SEGB12}) is equivalent to the field equation~(\ref{SEGB15}) for $\phi$ in the FLRW spacetime described by the metric~(\ref{FRW}). 

Let us consider the following example for a varying cosmological constant given by: 
\begin{align}
\label{SEGB17}
\Lambda(N) =&\, \frac{\Lambda_\mathrm{late} \e^{\left( N-N_\mathrm{EDE}\right)^{2n+1}}}
{\e^{\left(N - N_\mathrm{EDE}\right)^{2n+1}} + \frac{\sqrt{\Lambda_\mathrm{inf}\Lambda_\mathrm{late}} + \Lambda_\mathrm{late}}{\Lambda_\mathrm{EDE}}- 1} 
+ \frac{\Lambda_\mathrm{inf} \e^{- \left(N-N_\mathrm{EDE}\right)^{2n+1}}}
{\e^{ - \left( N - N_\mathrm{EDE}\right)^{2n+1}} + \frac{\sqrt{\Lambda_\mathrm{inf}\Lambda_\mathrm{late}} + \Lambda_\mathrm{inf}}{\Lambda_\mathrm{EDE}} - 1} \, 
\end{align}
Here $n$ is a non-negative integer, $n=0,1,2,3,\cdots$, and $\Lambda_\mathrm{inf}$, $\Lambda_\mathrm{EDE}$, and $\Lambda_\mathrm{late}$ are constants. 
We may assume that late-time including the present time corresponds to a large $N$. 
When $N$ is large, $\Lambda(N)$ goes to a constant $\Lambda_\mathrm{late}$, which we may be identified with the effective cosmological constant in the present universe. 
The limit $N\to - \infty$ corresponds to the early universe, when $\Lambda(N)$ goes to another constant $\Lambda_\mathrm{inf}$, 
which may be identified with the effective cosmological constant that produces inflation. 
On the other hand, when $N\sim N_\mathrm{EDE}$, we find $\Lambda(N)$ behaves also as a constant: 
\begin{align}
\label{SEGB18}
\Lambda(N) \sim \Lambda_\mathrm{EDE} + \mathcal{O}\left( \left(N-N_\mathrm{EDE}\right)^{4n+2} \right) \, .
\end{align}
Hence, $\Lambda_\mathrm{EDE}$ will be the effective cosmological constant corresponding to the early dark energy. 
In this scenario, the value of the effective cosmological constant in the early universe, at the moment of recombination approximately, was smaller than in the present universe. 
Therefore we find a hierarchy $0<\Lambda_\mathrm{EDE}<\Lambda_\mathrm{late}<\Lambda_\mathrm{inf}$. 
By contrast, in s single scalar field model, such unification might be impossible to be realized, since the early dark energy with 
a single scalar field $\Lambda_\mathrm{EDE}$ should correspond to a minimum of the potential of the scalar field if $\Lambda_\mathrm{EDE}$, $\Lambda_\mathrm{late}$, 
and $\Lambda_\mathrm{inf}$ correspond to the extrema of the potential. 
The minimum of the potential is stable and $\Lambda(N)=\Lambda_\mathrm{EDE}$ could not evolute to $\Lambda(N)=\Lambda_\mathrm{late}>\Lambda_\mathrm{EDE}$ 
because the scalar field would require climbing the potential. Nevertheless, in the case of scalar-Einstein-Gauss-Bonnet gravity, 
since $\Lambda(N) \neq V_\mathrm{eff}(\phi)$ but $\Lambda(N) = V(\phi)$, and the effective potential $V_\mathrm{eff}(\phi)$ depends on the curvature, 
such a transition is possible. 

\section{Inflation}\label{SecIII}

Let us now analyze the behavior of the model~(\ref{SEGB17}) during the inflationary epoch. 
For $N\to -\infty$, Eq.~(\ref{SEGB17}) can be expanded in a power series of $\e^{N-N_\mathrm{EDE}}$ as follows: 
\begin{align}
\label{inf1}
\Lambda(N) =&\, \Lambda_\mathrm{inf} \left[ 1 + \left\{ \frac{\Lambda_\mathrm{late}}
{\frac{\sqrt{\Lambda_\mathrm{inf}\Lambda_\mathrm{late}} + \Lambda_\mathrm{late}}{\Lambda_\mathrm{EDE}}- 1} 
 - \frac{\sqrt{\Lambda_\mathrm{inf}\Lambda_\mathrm{late}} + \Lambda_\mathrm{inf}}{\Lambda_\mathrm{EDE}} + 1
\right\} \e^{N - N_\mathrm{EDE}} 
+ \mathcal{O} \left( \left( \e^{N - N_\mathrm{EDE}} \right)^2 \right)
\right] \nonumber \\
\sim &\, \Lambda_\mathrm{inf} \left[ 1 - \frac{\Lambda_\mathrm{inf}}{\Lambda_\mathrm{EDE}} \e^{N - N_\mathrm{EDE}} 
+ \mathcal{O} \left( \left( \e^{N - N_\mathrm{EDE}} \right)^2 \right)
\right] \, .
\end{align}
For simplicity here, we only consider the case that $n=0$. Furthermore we have assumed $\Lambda_\mathrm{inf} \gg \Lambda_\mathrm{late} > \Lambda_\mathrm{EDE}$ and consequently 
$\frac{\Lambda_\mathrm{inf}}{\Lambda_\mathrm{EDE}} \e^{N - N_\mathrm{EDE}} \ll 1$. During inflation, any other matter contribution can be neglected, such that (\ref{GBEDE7}) becomes: 
\begin{align}
\label{inf3}
g(N) \sim \sqrt{\frac{\kappa^2 \Lambda_\mathrm{inf}}{3}} 
\left[ 1 - \frac{\Lambda_\mathrm{inf}}{2\Lambda_\mathrm{EDE}} \e^{N - N_\mathrm{EDE}} 
+ \mathcal{O} \left( \left( \e^{N - N_\mathrm{EDE}} \right)^2 \right)
\right] \, .
\end{align}
By using (\ref{phi2}), the behaviour of $\phi(N)$ during inflation is obtained: 
\begin{align}
\label{inf4}
\phi\left(N \right) = \pm \left(\frac{12}{\kappa^8 \Lambda_\mathrm{inf}}\right)^\frac{1}{4}N + \phi_0 
+ \mathcal{O} \left( \e^{N - N_\mathrm{EDE}} \right) \, .
\end{align}
And consequently, the function $f$ is given by: 
\begin{align}
\label{inf5}
N=f(\phi) \sim \pm \left(\frac{\kappa^2 \Lambda_\mathrm{inf}}{12}\right)^\frac{1}{4}\left( \phi - \phi_0\right) \, ,
\end{align}
where $\phi_0$ is a constant of the integration. 
Finally, the scalar potential and the coupling (\ref{GBEDE11}) are reconstructed: 
\begin{align}
\label{inf6}
\xi\left(\phi\right) =&\, \xi_0 + \left(\frac{3}{\kappa^4 \Lambda_\mathrm{inf}} \right)^\frac{3}{2} 
\left[ N + \frac{3\Lambda_\mathrm{inf}}{2\Lambda_\mathrm{EDE}} \e^{N - N_\mathrm{EDE}} 
+ \mathcal{O} \left( \left( \e^{N - N_\mathrm{EDE}} \right)^2 \right) \right] \nonumber \\
\sim &\, \left(\frac{3}{\kappa^4 \Lambda_\mathrm{inf}} \right)^\frac{3}{2} 
\left[ \pm \left(\frac{\kappa^2 \Lambda_\mathrm{inf}}{12}\right)^\frac{1}{4} \phi 
+ \frac{3\Lambda_\mathrm{inf}}{2\Lambda_\mathrm{EDE}} \e^{\pm \left(\frac{\kappa^8 \Lambda_\mathrm{inf}}{12}\right)^\frac{1}{4} \phi} \right] \, , \\
\label{inf7}
V\left(\phi\right) =&\, \Lambda_\mathrm{inf} \left[ 1 - \frac{7\Lambda_\mathrm{inf}}{6\Lambda_\mathrm{EDE}} \e^{N - N_\mathrm{EDE}} \right]
 -7 \sqrt{\frac{\Lambda_\mathrm{inf}}{3\kappa^4}} 
\left[ 1 - \frac{\Lambda_\mathrm{inf}}{2\Lambda_\mathrm{EDE}} \e^{N - N_\mathrm{EDE}} \right] 
+ \mathcal{O} \left( \left( \e^{N - N_\mathrm{EDE}} \right)^2 \right) \nonumber \\
\sim &\, -7 \sqrt{\frac{\Lambda_\mathrm{inf}}{3\kappa^4}} 
\left[ 1 - \frac{\Lambda_\mathrm{inf}}{2\Lambda_\mathrm{EDE}} \e^{\pm \left(\frac{\kappa^8 \Lambda_\mathrm{inf}}{12}\right)^\frac{1}{4} \phi} \right] \, .
\end{align}
Here $\xi_0$ is an integration constant and we have chosen $\xi_0= \left(\frac{3}{\kappa^4 \Lambda_\mathrm{inf}} \right)^\frac{3}{2} N_\mathrm{EDE}$. 
We have also chosen $\phi_0$ so that $\mp \left(\frac{\kappa^8 \Lambda_\mathrm{inf}}{12}\right)^\frac{1}{4} \phi_0 = N_\mathrm{EDE}$ 
and we have assumed $\frac{1}{\kappa^4}\gg \Lambda_\mathrm{inf}$. 
Under these assumptions, the effective potential $V_\mathrm{eff}(\phi)$ in (\ref{SEGB12_0}) yields:
\begin{align}
\label{inf9}
V_\mathrm{eff}(\phi) \sim&\, -7 \sqrt{\frac{\Lambda_\mathrm{inf}}{3\kappa^4}} 
\left[ 1 - \frac{\Lambda_\mathrm{inf}}{2\Lambda_\mathrm{EDE}} \e^{\pm \left(\frac{\kappa^8 \Lambda_\mathrm{inf}}{12}\right)^\frac{1}{4} \phi} \right] \nonumber \\
&\, + 4 \sqrt{\frac{\Lambda_\mathrm{inf}}{3\kappa^4}} \left[ \pm \left(\frac{\kappa^2 \Lambda_\mathrm{inf}}{12}\right)^\frac{1}{4} \phi 
+ \frac{3\Lambda_\mathrm{inf}}{2\Lambda_\mathrm{EDE}} \e^{\pm \left(\frac{\kappa^8 \Lambda_\mathrm{inf}}{12}\right)^\frac{1}{4} \phi} \right] 
\left[ 1 - \frac{4\Lambda_\mathrm{inf}}{\Lambda_\mathrm{EDE}} \e^{N - N_\mathrm{EDE}} \right] \nonumber \\
\sim&\, 4 \sqrt{\frac{\Lambda_\mathrm{inf}}{3\kappa^4}} \left(N-N_\mathrm{EDE} \right) \, ,
\end{align}
Moreover, the usual slow-roll parameters $\epsilon$ end $\eta$ are given by:
\begin{align}
\label{inf12}
\epsilon \equiv \frac{1}{4\kappa^2} \left( \frac{V_\mathrm{eff}'\left(\phi\left(N\right)\right)}{\Lambda(N)}\right)^2\, , \quad 
\eta \equiv \frac{V_\mathrm{eff}''\left(\phi\left(N\right)\right)}{2\kappa^2\Lambda(N)} \, ,
\end{align}
Note that the above slow-roll parameters are defined for the effective potential as given in the scalar field equation (\ref{SEGB12_0}), such that the slow-roll conditions $\epsilon<<1$ and $\eta<<1$ during inflation affects not only to the scalar field potential $V(\phi)$ but also the coupling $\xi(\phi)$. Other approaches define extra slow-roll parameters to impose such conditions for the realisation of slow-roll inflation on the coupling function separately (for more details, see Refs.~\cite{Oikonomou:2022xoq,Oikonomou:2021kql,Oikonomou:2015qha,Hwang:2005hb}). The derivatives of the effective potential (\ref{inf9}) yield:
\begin{align}
\label{inf10}
V_\mathrm{eff}'(\phi) 
\sim&\, \pm \left(\frac{\kappa^8 \Lambda_\mathrm{inf}}{12}\right)^\frac{1}{4} \sqrt{\frac{\Lambda_\mathrm{inf}}{3\kappa^4}} 
\left[ - \frac{7\Lambda_\mathrm{inf}}{2\Lambda_\mathrm{EDE}} \e^{\pm \left(\frac{\kappa^8 \Lambda_\mathrm{inf}}{12}\right)^\frac{1}{4} \phi} 
+ \left\{ 4 + \frac{12\Lambda_\mathrm{inf}}{2\Lambda_\mathrm{EDE}} \e^{\pm \left(\frac{\kappa^8 \Lambda_\mathrm{inf}}{12}\right)^\frac{1}{4} \phi} \right\}
\left\{ 1 - \frac{4\Lambda_\mathrm{inf}}{\Lambda_\mathrm{EDE}} \e^{N - N_\mathrm{EDE}} \right\} \right] \nonumber \\
\sim & \pm 4 \left(\frac{\kappa^8 \Lambda_\mathrm{inf}}{12}\right)^\frac{1}{4} \sqrt{\frac{\Lambda_\mathrm{inf}}{3\kappa^4}} \, , \\
\label{inf11}
V_\mathrm{eff}''(\phi) 
\sim&\, \left(\frac{\kappa^8 \Lambda_\mathrm{inf}}{12}\right)^\frac{1}{2} \sqrt{\frac{\Lambda_\mathrm{inf}}{3\kappa^4}} 
\left[ - \frac{7\Lambda_\mathrm{inf}}{2\Lambda_\mathrm{EDE}} \e^{\pm \left(\frac{\kappa^8 \Lambda_\mathrm{inf}}{12}\right)^\frac{1}{4} \phi} 
+ \frac{12\Lambda_\mathrm{inf}}{2\Lambda_\mathrm{EDE}} \e^{\pm \left(\frac{\kappa^8 \Lambda_\mathrm{inf}}{12}\right)^\frac{1}{4} \phi} 
\left\{ 1 - \frac{4\Lambda_\mathrm{inf}}{\Lambda_\mathrm{EDE}} \e^{N - N_\mathrm{EDE}} \right\} \right] \nonumber \\
\sim&\, \left(\frac{\kappa^8 \Lambda_\mathrm{inf}}{12}\right)^\frac{1}{2} \sqrt{\frac{\Lambda_\mathrm{inf}}{3\kappa^4}} 
\frac{5\Lambda_\mathrm{inf} \e^{N-N_\mathrm{EDE}}}{2\Lambda_\mathrm{EDE}} \, .
\end{align}
And the slow-roll parameters read as follows:
\begin{align}
\label{inf14}
\epsilon \sim \frac{1}{3\kappa^2\sqrt{\Lambda_\mathrm{inf}}}\sqrt{\frac{2}{3}} \, , \quad \eta\sim 0\, ,
\end{align}
Hence, the spectral index $n_s$ and the tensor-to-scalar ratio $r$ can be obtained:
\begin{align}
\label{inf13}
n_s - 1 = - 6\epsilon + 2 \eta\, , \quad 
r=16\epsilon\, ,
\end{align}
which in this case leads to:
\begin{align}
\label{inf13}
n_s - 1 \sim - \frac{2}{3\kappa^2\sqrt{\Lambda_\mathrm{inf}}}\sqrt{\frac{2}{3}}\, , \quad 
r \sim \frac{16}{3\kappa^2\sqrt{\Lambda_\mathrm{inf}}}\sqrt{\frac{2}{3}}\, .
\end{align}
One can note that both parameters depend on the scale of inflation and consequently, this scale can be set appropriately in order to fit 
the constraints provided by the Planck data~\cite{Planck:2018vyg}.

\section{Phantom dark energy}\label{SecIV}

In this section, we consider the possible scenario of phantom dark energy instead of a varying cosmological constant considered above. 
Phantom dark energy is usually described by a perfect fluid whose EoS parameter $w_\mathrm{phantom}<-1$, 
such that the energy density of the phantom fluid increases with the expansion of the universe: 
\begin{align}
\label{phantom1}
\rho_\mathrm{phantom} = \rho_{\mathrm{phantom}\,0} \e^{-3 \left( 1 + w_\mathrm{phantom} \right) N}\, , 
\end{align}
where $-3 \left( 1 + w_\mathrm{phantom} \right)>0$. 
Since the energy density increases along the expansion, the Hubble parameter measured today $H_0$ might be larger than the Hubble rate 
at the time of recombination and consequently, the phantom scenario might provide predictions consistent with the observations and particularly with the tension on the Hubble rate.

The reconstruction of a phantom fluid is not possible in standard scalar-tensor theory without any Gauss-Bonnet terms, where the action for a canonical scalar field is given by:
\begin{align}
\label{phantom2}
S=\int d^4 x \sqrt{-g}\left[ \frac{R}{2\kappa^2} 
 - \frac{1}{2}\partial_\mu \phi \partial^\mu \phi - V(\phi) \right]\, . 
\end{align} 
In a FLRW universe (\ref{FRW}), the energy density $\rho_\phi$ and the pressure $p_\phi$ for the scalar field $\phi$ are given by:
\begin{align}
\label{phantom3}
\rho_\phi = \frac{1}{2}{\dot\phi}^2 + V(\phi)\, , \quad 
p_\phi = \frac{1}{2}{\dot\phi}^2 - V(\phi)\, .
\end{align}
Hence, since the energy density is defined as positive, the potential is non-negative $V(\phi)\geq 0$, and consequently the EoS parameter $w_\phi$ is always greater than $-1$,
\begin{align}
\label{phantom4}
w_\phi \equiv \frac{p_\phi}{\rho_\phi} = -1 + \frac{{\dot\phi}^2}{\frac{1}{2}{\dot\phi}^2 + V(\phi)} > -1 \, ,
\end{align}
Nevertheless, one might construct a phantom dark energy model within the framework of scalar-tensor theory by assuming a negative kinetic term, 
\begin{align}
\label{phantom5}
S=\int d^4 x \sqrt{-g}\left[ \frac{R}{2\kappa^2} 
+ \frac{1}{2}\partial_\mu \phi \partial^\mu \phi - V(\phi) \right]\, ,
\end{align} 
whose energy density $\rho_\phi$ and the pressure $p_\phi$ are given by 
\begin{align}
\label{phantom6}
\rho_\phi = - \frac{1}{2}{\dot\phi}^2 + V(\phi)\, , \quad 
p_\phi = - \frac{1}{2}{\dot\phi}^2 - V(\phi)\, , 
\end{align}
which gives the following EoS parameter, 
\begin{align}
\label{phantom7}
w_\phi \equiv \frac{p_\phi}{\rho_\phi} = -1 - \frac{{\dot\phi}^2}{- \frac{1}{2}{\dot\phi}^2 + V(\phi)} \, ,
\end{align}
and therefore $w_\phi$ can be smaller than $-1$. 
However, the model (\ref{phantom5}) is physically inconsistent because from the classical point of view, 
the energy density $\rho_\phi$ is unbounded from below, and from the quantum theory, the scalar field becomes a ghost, 
which generates negative norm states and therefore the model contradicts the basic concept of quantum mechanics.

In~\cite{Gangopadhyay:2022bsh}, a scenario to reproduce phantom dark energy without the presence of ghosts has been proposed 
by assuming a coupling between baryonic matter and dark matter that shows a phantom transition. 
Also in the framework of the scalar-Gauss-Bonnet gravity, a phantom universe can be realized without ghosts~\cite{Nojiri:2005vv}. 
Hence, here we are considering the possibility of reproducing phantom dark energy within the scalar-Einstein-Gauss-Bonnet gravity~(\ref{SEGB1}). 
To do so, since the effective cosmological constant increases with time in the phantom scenario, 
we may simply consider the model for $\Lambda(N)$ in Eq.~(\ref{GBEDE7}) given by:
\begin{align}
\label{SEGB17ph}
\Lambda(N) = \Lambda_0 \e^{\alpha N}\, .
\end{align}
Here $\Lambda_0$ and $\alpha$ are positive constants. 
As $\Lambda(N)$ increases with the number of $e$-foldings $N$, Eq.~(\ref{SEGB17ph}) provides the total energy density asymptotically, 
such that the effective EoS parameter $w$ leads to: 
\begin{align}
\label{mphantom}
w = - 1 - \frac{\alpha}{3}\, .
\end{align}
And a phantom universe occurs for $\alpha>0$. 
Hence, for this model, the corresponding Hubble parameter (\ref{GBEDE7_0}) is given by:
\begin{align}
\label{pexp1}
H(N)^2 = \frac{\kappa^2}{3} \left( \Lambda_0 \e^{\alpha N} + \rho_0\e^{-3N} \right)\, .
\end{align}
Here we have neglected every other contribution except a pressureless fluid that accounts for dark matter and baryons. 
By assuming $N=0$ corresponding to the present time, $\rho_0$ is the sum of the energy densities of dark matter and ordinary matter today 
and by assuming the value of the Hubble constant as provided by the Planck collaboration, $H_0 = 67.36 \pm 0.54$ km/s/Mpc, 
one finds $\Lambda_0 + \rho_0=8.15\times 10^{-27} \mathrm{kg\ m^{-3}}$, and since dark energy represents 68.3\% of 
the total energy density, $\Lambda_0 =5.56\times 10^{-27} \mathrm{kg\ m^{-3}}$. 
Nevertheless, the cosmological constant varies with time within the model (\ref{SEGB17ph}), such that at the epoch of recombination, 
which corresponds to $\e^{N_\mathrm{CMB}}\sim 1/1100$, the corresponding cosmological constant according to the Planck data is:
\begin{align}
\label{pexp4}
\Lambda_0 \e^{\alpha N_\mathrm{CMB}}=5.56\times 10^{-27} \mathrm{kg\ m^{-3}}\, ,
\end{align} 
where the number of $e$-foldings at the time of recombination is:
\begin{align}
\label{pexp3}
N_\mathrm{CMB} \sim - 7.00\, . 
\end{align}
On the other hand, low-$z$ observations provide a value of the Hubble rate greater than the one from Planck, $H_0 \sim 73$-$74 $ km/s/Mpc, 
which leads to $\Lambda_0 + \rho_0=9.70\times 10^{-27} \mathrm{kg\ m^{-3}}$ and the present dark energy density is given by: 
\begin{align}
\label{pexp5}
\Lambda_0 =6.63\times 10^{-27} \mathrm{kg\ m^{-3}}\, .
\end{align}
By comparing (\ref{pexp4}) and (\ref{pexp5}), and using (\ref{pexp3}), one finds: 
\begin{align}
\label{pexp6}
\alpha=0.0251\, .
\end{align}
Then, we can check and compare with the observed value the time at which the transition from decelerating expansion to accelerating expansion occurs, which corresponds to: 
\begin{align}
\label{pexp7}
0 = \frac{\ddot a}{a}=\dot H + H^2 = H' H + H^2\, .
\end{align}
By using (\ref{pexp1}), we obtain, 
\begin{align}
\label{pexp8}
0= \left( 1 + \frac{\alpha}{2} \right) \Lambda_0 \e^{\alpha N_\mathrm{tra}} - \frac{\rho_0}{2} \e^{-3N_\mathrm{tra}} \, .
\end{align}
Here $N_\mathrm{tra}$ is the number of $e$-foldings corresponding to the transition from decelerating expansion to accelerating expansion 
evaluated in terms of the present scale factor. 
By using (\ref{pexp8}), the corresponding value leads to:
\begin{align}
\label{pexp9}
N_\mathrm{tra}=-0.49 \, .
\end{align}
Observations of Supernovae point that the transition occurred at redshift $z=0.67$, which in terms of the number of $e$-foldings, 
$N=-\ln \left( 1 + z \right)$, gives $N_\mathrm{tra}=-0.51$, which is compatible with (\ref{pexp9}), as predicted by this model. 
This is possible since $\alpha$ in (\ref{pexp6}) is small enough. 
Hence, the model (\ref{SEGB17ph}) might be consistent with observations, providing at the same time a cosmological constant that evolutes with the expansion, 
such that its different values at the present epoch in comparison to the epoch of recombination may shed some light on the tension 
of the Hubble constant as provided by early and late observational data. 

\section{Some generalizations of EGB gravity}\label{SecV}

Let us now consider a generalization of the action for EGB gravity. 
The simplest way of generalizing the gravitational action (\ref{SEGB1}) is to promote the Ricci scalar term 
to an arbitrary function coupled to the scalar field~\cite{Odintsov:2022zrj}: 
\begin{align}
\label{SEGB19}
S=\int d^4 x \sqrt{-g}\left[ \frac{\eta\left(R, \phi \right)}{2\kappa^2} 
 - \frac{1}{2}\partial_\mu \phi \partial^\mu \phi 
 - V(\phi) - \xi(\phi) \mathcal{G}\right]\, ,
\end{align}
where $\eta\left(R.\phi \right)$ is a function of the scalar curvature $R$ and the scalar field $\phi$. 
In comparison to EGB gravity as described by the action (\ref{SEGB1}), this type of model includes ghost modes in general, 
such that the model might not be considered a consistent effective theory. 
Nevertheless, by considering the particular case where $\eta\left(R, \phi \right)=\zeta(\phi)R$, i.e., a Brans-Dicke-like model, ghost instabilities might be avoided. 
The field equations are obtained by varying the action~(\ref{SEGB19}) with respect to the scalar field $\phi$ and the metric $g_{\mu\nu}$, which lead to:
\begin{align}
\label{SEGB20}
0=&\, \frac{\zeta'\left(\phi \right) R}{2\kappa^2} + \Box \phi - V'(\phi) - \xi'(\phi) \mathcal{G}\, , \\
\label{SEGB21}
T^{\mu \nu}= &\, \frac{\zeta\left( \phi \right)}{2\kappa^2}\left(- R^{\mu\nu} + \frac{1}{2} g^{\mu\nu} R\right) 
+ \frac{1}{2\kappa^2} \left( \nabla^\mu \nabla^\nu - g^{\mu\nu} \Box \right) \zeta\left( \phi \right)
+ \left(\frac{1}{2} \partial^\mu \phi \partial^\nu \phi
 - \frac{1}{4}g^{\mu\nu} \partial_\rho \phi \partial^\rho \phi \right) - \frac{1}{2} g^{\mu\nu} V(\phi) \nonumber \\
&\, + 2 \left( \nabla^\mu \nabla^\nu \xi(\phi)\right)R - 2 g^{\mu\nu} \left( \nabla^2\xi(\phi)\right)R
 - 4 \left( \nabla_\rho \nabla^\mu \xi(\phi)\right)R^{\nu\rho}
 - 4 \left( \nabla_\rho \nabla^\nu \xi(\phi)\right)R^{\mu\rho} \nonumber \\
&\, + 4 \left( \nabla^2 \xi(\phi) \right)R^{\mu\nu}
+ 4g^{\mu\nu} \left( \nabla_{\rho} \nabla_\sigma \xi(\phi) \right) R^{\rho\sigma} 
 - 4 \left(\nabla_\rho \nabla_\sigma \xi(\phi) \right) R^{\mu\rho\nu\sigma}\, .
\end{align}
For a flat FLRW spacetime (\ref{FRW}), the equations turn out: 
\begin{align}
\label{SEGB22}
0=&\, - \frac{3\zeta(\phi) }{\kappa^2}H^2 - \frac{3}{2\kappa^2} H^2 \frac{d\zeta\left(\phi\left(N\right)\right)}{dN}
+ \frac{1}{2}H^2{\phi'}^2 + V(\phi) + 24 H^4 \frac{d \xi(\phi(N))}{dN} 
+ \sum_m \rho_{m0} \e^{-3 \left( 1 + w_m \right) N} \, ,\nonumber \\
0 =&\, \frac{\zeta(\phi)}{\kappa^2}\left(2H H' + 3 H^2 \right) + \frac{1}{2\kappa^2} \left\{ H^2 \frac{d^2\zeta\left(\phi\left(N\right)\right)}{dN^2} 
+ \left( HH' + 2 H^2 \right) \frac{d\zeta\left(\phi\left(N\right)\right)}{dN} \right\}
+ \frac{1}{2}H^2 {\phi'}^2 - V(\phi) \nonumber \\
&\, - 8 \left( 2 H^4 \frac{d \xi (\phi(N))}{dN} + 3 H^3 H' \frac{d \xi(\phi(N))}{dN} + H^4 \frac{d^2 \xi(\phi(N))}{dN^2} \right) 
+ \sum_m w_m \rho_{m0} \e^{-3 \left( 1 + w_m \right) N} \, ,
\end{align}
whereas in terms of the number of $e$-foldings, they yield: 
\begin{align}
\label{GBEDE23}
0=&\, \frac{2\zeta(\phi) H H'}{\kappa^2} + \frac{1}{2\kappa^2} \left\{ H^2 \frac{d^2\zeta\left(\phi\left(N\right)\right)}{dN^2} 
+ \left( HH' - H^2 \right) \frac{d\zeta\left(\phi\left(N\right)\right)}{dN} \right\} + H^2 {\phi'}^2 \nonumber \\
&\, - 8 H \e^N \frac{d}{dN} \left( \e^{-N} H^3 \frac{d \xi (\phi(N))}{dN} \right) + \sum_m \left( 1 + w_m\right) \rho_{m0} \e^{-3 \left( 1 + w_m \right) N} \, ,
\end{align}
By following the same procedure as described in Section~\ref{SecII}, by considering the following potential and coupling with the GB term: 
\begin{align}
\label{GBEDE24}
\xi\left(\phi\right) =&\, \frac{1}{8} \int_{N^{(2)}}^{f(\phi)} dN_1\frac{\e^{N_1}}{g \left( N_1 \right)^3} \int_{N^{(1)}}^{N_1} dN_2 \e^{-N_2} 
\left[ \frac{2\zeta \left(f^{-1}\left(N_2\right) \right) g' \left(N_2 \right)}{\kappa^2} 
+ \frac{1}{2\kappa^2} \left\{ g \left( N_2 \right) \frac{d^2\zeta \left(f^{-1}\left(N_2\right) \right)}{{dN_2}^2} \right. \right. \nonumber \\
&\, \left. \left. + \left( g' \left( N_2 \right) - g \left( N_2 \right) \right) \frac{d\zeta \left(f^{-1}\left(N_2\right) \right) }{dN_2} \right\} 
+ \frac{g \left( N_2 \right)}{f' \left( f^{-1}\left(N_2\right) \right)^2} 
+ \frac{1}{g\left(N_2\right)} \sum_m \left( 1 + w_m\right) \rho_{m0} \e^{-3 \left( 1 + w_m \right) N_2} \right]\, , \\
\label{GBEDE25}
V\left(\phi\right) =&\, \frac{3\zeta(\phi)g\left(f\left(\phi\right)\right)^2}{\kappa^2} 
+ \frac{3H^2 \zeta'(\phi)}{2\kappa^2f'\left(f^{-1}\left(\phi\right)\right)} - \frac{g\left(f\left(\phi\right)\right)^2}{2f'\left(\phi \right)^2} 
 - 3 g\left(f\left(\phi\right)\right) \e^{f(\phi)} \int_{N^{(1)}}^{f(\phi)} dN_1 \e^{-N_1} \nonumber \\
&\, \times \left[ \frac{2 \zeta \left(f^{-1}\left(N_1\right) \right) g' \left(N_1 \right)}{\kappa^2} 
+ \frac{1}{2\kappa^2} \left\{ g \left( N_1 \right) \frac{d^2\zeta \left(f^{-1}\left(N_1\right) \right)}{{dN_1}^2} 
+ \left( g' \left( N_1 \right) - g \left( N_1 \right) \right) \frac{d\zeta \left(f^{-1}\left(N_1\right) \right) }{dN_1} \right\} \right. \nonumber \\
&\, \left. \quad + \frac{g\left( N_1 \right)}{f' \left( f^{-1}\left(N_1\right) \right)^2} 
+ \frac{1}{g\left(N_1\right)} \sum_m \left( 1 + w_m\right) \rho_{m0} \e^{-3 \left( 1 + w_m \right) N_1} \right] 
 - \sum_m \rho_{m0} \e^{-3 \left( 1 + w_m \right) f^{-1}(\phi)} \, , 
\end{align}
the solution is given by (\ref{GBEDE6}). 
Note that now there is an extra ingredient in the model, given by the coupling to the Ricci scalar $\zeta(\phi)$, which remains arbitrary, 
such that might be chosen to simplify the equations. 
Let us consider the case given by~\cite{Odintsov:2022zrj}: 
\begin{align}
\label{GBEDE26}
\zeta(\phi)=\frac{\phi_0}{\phi}\, , \quad V(\phi)=\frac{V_0\phi_0}{\phi}\, , \quad \dot\xi=Ca \, ,
\end{align}
where $\phi_0$, $V_0$, and $C$ are constants. 
Such a model is to be compatible with observations~\cite{Odintsov:2022zrj}. 
Then, we might consider the Hubble parameter given in~(\ref{GBEDE7}) and (\ref{SEGB17}). 
By following the same procedure as in the previous section the value of $\Lambda_\mathrm{EDE}$ can be set as:
\begin{align}
\label{GBEDE27}
\Lambda_\mathrm{EDE}=5.56\times 10^{-27} \mathrm{kg\ m^{-3}}\, .
\end{align}
By assuming $N_0$ to be the number of $e$-foldings at the present time, Eq.~(\ref{pexp3}) provides $\e^{N_0-N_\mathrm{EDE}}\sim 1,100\gg 1$ 
while the expression for the varying cosmological constant~(\ref{SEGB17}) imposes: 
\begin{align}
\label{GBEDE28}
\Lambda \left( N_0 \right) \sim \Lambda_\mathrm{late} \, .
\end{align}
We might compare (\ref{GBEDE28}) with (\ref{pexp5}), such that:
\begin{align}
\label{GBEDE29}
\Lambda_\mathrm{late} =6.63\times 10^{-27} \mathrm{kg\ m^{-3}}\, .
\end{align}
Hence, the gravitational action (\ref{SEGB19}) with (\ref{GBEDE7}) and (\ref{SEGB17}) also unifies the cosmological evolution under the same mechanism, 
that is a varying cosmological constant produced by the evolution of a scalar field non-minimally coupled.

\section{Conclusion}

In summary, we have reconstructed a model that might explain the inflationary paradigm during the very early universe, the accelerating expansion of the present universe, 
and the early dark energy in a unified way by using the reconstructing the corresponding gravitational action in the framework of scalar-Einstein-Gauss-Bonnet gravity. 
The model reproduces inflation successfully and describes realistic dark energy simultaneously. In addition, the model may solve the Hubble tension problem by inducing an early dark energy without breaking the reasonable properties of inflation and the present dark energy epochs.  All rely on an effective varying constant that is produced by the effects of a scalar field coupled to the Gauss-Bonnet invariant in the action. Such cosmological constant may reconcile the values provided by the Planck collaboration and the one obtained from low-redshifts observations as Sne Ia/Cepheids, since the cosmological constant grows with the expansion, such that both datasets might provide correct measurements of the Hubble constant, 
the failure lies on considering the $\Lambda$CDM model as the fiducial model, as performed with the data from the CMB, where the Hubble constant is estimated 
by using the luminosity distance and the positions and heights of the CMB power spectrum peaks based on the $\Lambda$CDM model. 

The Hubble constant given by $\Lambda_\mathrm{EDE}$ in~(\ref{SEGB17}), for example, might be different from one estimated by Planck collaboration, since it differs from the current one, as it evolves with the expansion, transiting from one stage to the next. In addition, we have shown that inflation can be well realised in this framework, where the corresponding spectral index and tensor-to-scalar ratio  can fit well the values provided by CMB observations by setting the appropriate value of a single parameter. Moreover, the model shows that a phantom transition might occur without the usual instabilities of standard scalar fields. 

However, one might point that in our formalism the presence of extra degrees of freedom allows the reconstruction of any expansion of the universe which can be modified appropriately to fit the observational data. Nevertheless, the model tries to unify different behaviours of the cosmological expansion along the whole cosmological history, such that whether one compares this unified perspective from the usual models for inflation and (early) dark energy, there are not actually extra degrees of freedom. 

Moreover, this analysis should work out as an excellent starting point to go further in the analysis of scalar-Einstein-Gauss-Bonnet gravity, since it provides a correct description of the universe expansion, sheds some light about the possible solution to the Hubble tension problem and at the same time is absent of ghosts. Further analysis focused on other gravitational aspects should show the definite test for this type of theories.

Hence, in this paper we have proposed a way to construct solutions within the scalar-Einstein-Gauss-Bonnet gravity and showed that a unified description of the whole cosmological history can be achieved, including a possible solution to the Hubble tension problem.

\section*{Acknowledgements}
This work is supported by the Spanish National Grants PID2019-104397GB-I00 (S. D. O.) and PID2020-117301GA-I00 (D.SC.G.) funded by MCIN/AEI/10.13039/501100011033 (``ERDF A way of making Europe" and ``PGC Generaci\'on de Conocimiento").S. D. O. was also supported by JSPS visiting fellowship S23013. D.SC.G. also acknowledges the hospitality of the Institute of Space Sciences (ICE-CSIC) for a visit during which part of this work was carried out.

\end{document}